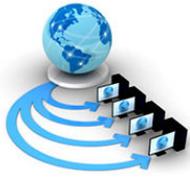

# Evaluation of Fiji National University Campus Information Systems

Bimal Aklesh Kumar
Department of Computer Science and Information Systems
Fiji National University
Fiji Islands
bimal.kumar@fnu.ac.fj

*Abstract:* Fiji National University (FNU) has been encountering many difficulties with its current campus administrative systems. These difficulties include accessibility, scalability, performance, flexibility and integration. In order to address these difficulties, we developed a thin client web based campus information system. The newly designed system allows the students, academic and administration staff of the university to handle their day to day affairs with the university online. In this paper we describe three types of evaluation carried out to determine the suitability of newly developed system for FNU environment. User interface evaluation was carried out to assess user interface on a set of usability principles, usability evaluation to see the ease at which users can use the system and finally performance evaluation to verify and validate user response time required to complete various tasks. The result of each of these evaluations were analysed and the system was rectified as part of iterative design process.

## I. INTRODUCTION

Fiji National University (FNU) was established in 2010 with the merger of six government owned tertiary institutions. It is a national institution, supporting the national effort for a stable economy and a literate population that is able to establish itself in the global community, while understanding and responding to the aspirations of individuals. FNU has a network of thirteen campuses throughout the country. The objective of the FNU is to promote research and academic excellence for the welfare and needs of the communities in Fiji as well as communities in the region and abroad who wish to receive tertiary education of high quality at affordable cost [3].

Prior to the merger and formation of the FNU and due to the autonomous operations of these colleges, at least three different campus information systems existed [4]. The university faced considerable amount of difficulties with these systems. These difficulties included accessibility, scalability, performance, flexibility and integration. In order to address these difficulties, we developed thin client web-based campus information systems. It was built using open source products and tools on modern code base with modern databases. FNU-CIS has relatively clean separation between presentation, business logic, and data access layers, with solid data architectures and a well defined set of business processes. It is easily accessible to all the students and the staff of FNU through the local intranet or via World Wide Web. The design is such where subsequent modification is limited as possible to least cost effect components and would not result in chain reaction of compensating modification hence making it easier to add more functionality in future [3]. The system easily integrates with other systems such as finance and human resource used by the university.

In order to assess the suitability of newly developed system for FNU we believe that three types of evaluation would be required; user interface evaluation to assess the user interface based on a set of usability principles, usability evaluation to see the ease at which users can use the system and finally performance evaluation to verify the turnaround time required to complete various tasks. This paper describes in detail three types of evaluation carried out and presents the results of this evaluation.

## II. RELATED WORK

In literature there are very few studies reporting evaluation of information systems used by universities. Whyte and Bythway (1996) proposed a holistic approach to IS evaluation by specifying three core elements to a system: Product that is hardware, software and training provided to the users; service that is how users are responded to and process by which product and services are provided. Gemmell and Pagano (2003) used product-service-process grid to analyze the student information systems at Salford University, UK. The attributes associated with each element was then evaluated by the users and finally gap approach was taken to the measurement of those system attributes (importance and performance). It has been recognized in literature that user satisfaction significantly affects the success or failure of any information systems [2]. Davis (2006), According to the technology acceptance model (TAM), simplicity, perceived ease of use and efficiency are three fundamental determinants of user satisfaction. Therefore, we believe that user interface evaluation, usability evaluation and performance evaluation would help as measure user satisfaction and success of our newly developed system. This paper advances the body of software evaluation knowledge at higher education sector.

## III. USER INTERFACE EVALUATION

There are basically four ways to evaluate a user interface: *formally* by some analysis technique, *automatically* by a computerized procedure, *empirically* by experiments with test users, and *heuristically* by simply looking at the interface and passing judgment according to ones own opinion. Formally analysis models are currently under extensive research but they have not reached the stage where they can be generally applied in real software development projects. Automatic evaluation is completely infeasible excerpt for a very few primitive checks. Therefore current practice is to do empirical evaluations if one wants a good and thorough evaluation of a user interface.

Unfortunately in most practical cases people actually do not conduct empirical evaluations because they lack time, expertise, inclination or simply the tradition to do so. In real life most user interface evaluations are heuristic evaluations [8].

Heuristic evaluation is usability engineering method for finding the usability problems in a user interface design so that they can be attended to as part of iterative design process [8]. The process requires that a small set of testers (evaluators) examine the interface and judge its compliance with recognized usability principles (heuristics) [13]. The goal is the identification of any usability issues, so that they can be addressed to as part of the iterative design process. Heuristic evaluation is popular in web development circles because it requires few resources in terms of money, time or expertise, so any developer can enjoy the following benefits of heuristic evaluation [12].

- This method provides quick and relatively cheap feed back for designers, the results would generate good ideas for improving user interfaces. The development team will also receive a good estimate how much the user interface can be improved.
- There is general acceptance that the design feedback provided by the method is valid and useful. It can also be obtained early on in the design process, whilst checking conformity to established guidelines helps promote compatibility with other systems.
- This method can seem overly critical as designers may only get a feed back on the problematic aspects of the interface as the method is normally used for the identification of good aspects.
- Usability problems found are normally restricted to aspects of the interface that are reasonably easy to demonstrate: use of colors, lay-out and information structuring, consistency of the terminology, consistency of the interaction mechanism.

There are three phases to carrying out heuristic evaluation, planning, running and report [14]. In planning a panel of experts is established with materials and equipments for evaluations. The experts work with the system while a list of problems and recommendations are created in the third.

### A. Planning

The panel of experts must be established in good time for the evaluation. The materials and the equipment for the demonstration should also be in place. All the analysts need to have sufficient time to become familiar with the product along with the intended task scenarios. They should operate by an agreed set of evaluation criteria [8]. Our system would have three primary users' students, academic and administration staff. We chose a set of five evaluators from each group to examine the interface and judge its compliance with a set of recognized usability principles given on table I. These heuristics are general rules that described the common properties of usable interfaces.

Table I. Usabilty Principles

| No. | Heuristics |
|---|---|
| 1 | Simple and natural dialogue |
| 2 | Speak the users language |
| 3 | Minimize user memory Load |
| 4 | Be consistent |
| 5 | Provide feedback |
| 6 | Provide clearly marked exists |
| 7 | Provide shortcuts |
| 8 | Give Error messages |
| 10 | Help and documentation |

System check list was produced based on the above heuristics for evaluators to use as a guide. Evaluators were required to identify problems and provide recommendations based on the severity ratings. Severity rating is allocated to each problem which indicates the most serious problems. The following 5 scale severity given on table II was used.

Table II. Severity Ratings

| Scale | Description |
|---|---|
| 0 | I don't agree that this is a usability problem at all |
| 1 | Cosmetic problem only: need not be fixed unless extra time is available on project |
| 2 | Minor usability problem: fixing this should be given low priority |
| 3 | Major usability problem: important to fix, so should be given high priority |
| 4 | Usability catastrophe: imperative to fix this before product can be released |

### B. Running

The experts should be aware of any relevant contextual information relating to the intended user group, tasks and usage of the product. Heuristic evaluation is performed by having each evaluator inspect the interface alone. Only after all evaluations have been completed are the evaluators allowed to communicate and have their findings aggregated, this procedure is important in order to ensure independent and unbiased evaluations from each evaluator [14]. The results of the evaluation can be recorded as written reports these reports have the advantage of presenting a formal record of evaluation but require additional effort by the evaluator and to read and aggregated by the evaluation manager. During the evaluation session, the evaluators went through the interfaces several times and inspected the various elements and compared them with the list of heuristics.

### C. Report

A list of identified problems which may be prioritized with regards to the severity rating and safety critical is produced. A report detailing the identified problems is written and provided as feedback to the development team. Heuristic evaluation does not provide a systematic way to generate fixes to the usability problems or a way to assess the probable quality of any redesigns [15]. However because heuristic evaluation aims at explaining each observed usability problem with reference to established usability principles, it will often be fairly easy to generate a revised design according to guidelines provided by the violated principle for good interactive systems [13]. Also many usability problems have fairly obvious fixes once they have been identified. There were four problems with a severity of three and above which is of high priority and it is important to fix. Other problems had a severity rating of one and two. These were minor problems. Table III, discusses the recommendation of the problems with severity rating of three and all of the problems found by the evaluators were implemented.

Table III. Recommendations

| Problem No. | Recommendation |
|---|---|
| 2 | Provide explanation on technical jargons used. |
| 5 | Display appropriate messages while processing is taking place in the background. |
| 8 | Fields should be checked before passing information to database. Apply error checking of fields and if there is an error display a pertinent explanation. |
| 10 | Implement the help feature |

## IV. USABILITY EVALUATION

Usability testing requires number of users to perform a set of pre-defined tasks [5]. During the testing evaluators assess how the users interact with the system and identify the usability issues of the system. Usability is one of the most important success factors in system quality in particular for web sites. Testing the web usability appears to be more difficult than testing traditional systems [7]. This is for the two reasons firstly the web users are located all over the world but they access it concurrently and secondly different types of hardware and software are used in order to access the web. The usability of web-based systems has a great impact on users on a daily basis, users are unlikely to revisit a site if they encounter difficulties in using the site, where alternative sites are available.

The limitations to usability testing are; firstly that testing is always an artificial situation which lack realistic circumstances and secondly participants do not fully represent the targeted web site audience [10]. There are various methods for usability evaluation. Model/Metrics based use model of tool to generate usability measure. Inquiry based communicates with users to gain insights into usability problems. Inspection method reviews the user interface and tries it out to find problems. Testing method collects data to be analyzed while a user uses the system [7]. In our study we believe that testing method would be appropriate. The following steps are required to carry out evaluation [10]:

- Test Plan - document what you will test and how. Include major goals of the test.
- Product/Prototype - the product must be bug free as appropriate for the goals for test. Pilot test the prototype by trying out the test scenarios. Consider all parts of the system.
- Experimental Design - basic usability test of a single product usually an informal design. Competitive or comparative studies.
- Test Participants - participants to the level of the development. Number of participants depends on the level of the test.
- Scenarios - these are usually end user "use cases" for testing the product.
- Test Setup - fidelity of test setup based on goals of test. Common laboratory set up.
- Test Procedures - typical test consider limits of attention in test time
- Measures - perform measures. Secondary task performance
- Analysis - statistical comparison of competitive tasks.

For the purpose of usability evaluation we did a comparative study by evaluating user satisfaction, working with the various features of FNU-CIS and one of the existing campus management systems PREMIUM. Usability is not a single one dimensional property; it has multiple components with various attributes associated with user interface. These measures are the standard ones for determining how "usable" an interactive the system is, and allows us to make judgments on the suitability of the interface for the tasks being carried out. Efficiency was measured in terms of ease to use the system. Errors are any action that prevents successful occurrence of desired result and since some errors escalate the users' time, its effect is measured in the efficiency of use. Learn-ability and satisfaction was a subjective measure assigned by each participant in the experiment. Interface memorability is rarely tested as thoroughly as other attributes but having the comparison and post test questionnaires of both systems made it feasible to some extent.

Set of task list was prepared for the evaluators to work on both the systems. Participants with an equal mixture of students, academic and administration staff, volunteered for our usability study of FNU-CIS and Premium System. We identified 45 participants to carry out a set of task on two systems. After completion of the assigned tasks, participants answered the post-test questionnaire and ranked the systems in order of preference. Every participant undertook all these tasks according to the task lists and used the two systems across a number of sessions. After completing tasks, participants were asked to fill out a post-test questionnaire that contained a subjective rating assigned to each tested characteristic of the system.

The post-test questionnaire consisted of a 5-point rating scale to gauze each characteristic of both applications. The rating scale ranged from 1 to 5 where 1 is "Strongly Disagree" and 5 is "Strongly Agree". There were also open ended questions to gain user feedback. The tested features include: login, enrollment, generating class list, course adjustments, retrieving course work, accessing student details and student grades. The bar chart in figure 1 presents the average ratings for the tested features on efficiency like wise bar chart on figure 2 presents the average rating on preference.

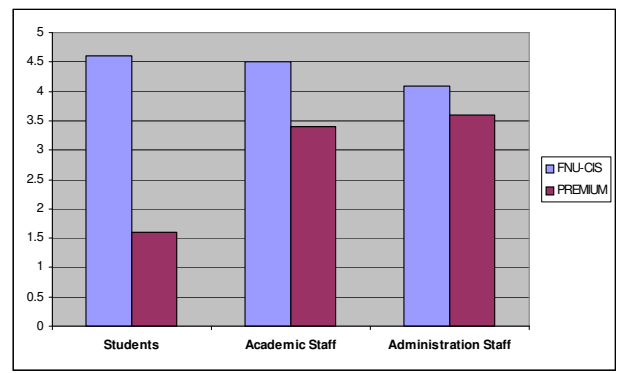

Figure 1. Usabilty test results on effeciency.

All three groups of participants found that FNU-CIS is more efficient than Premium system.

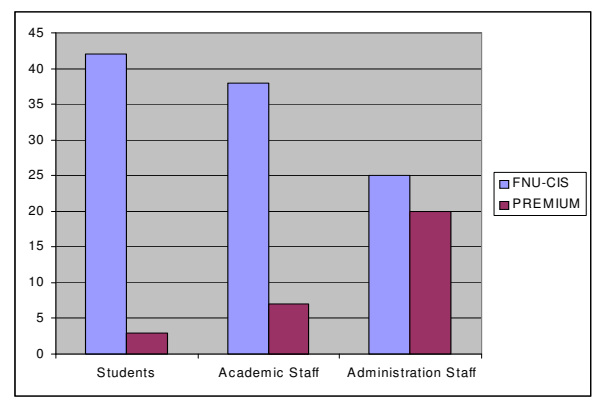

Figure 2. Usabilty test results on preference.

All three groups of participants favored to use FNU-CIS over Premium system.

     3

## V. PERFORMANCE EVALUATION

Performance evaluation determines how fast some aspects of a system perform under a particular work load. It can also serve to validate and verify other quality attributes of the system [11]. Performance evaluation can serve different purposes. It can demonstrate that the system meets performance criteria, can be used to compare two systems to find out which one performs better or measure what parts of the system performs badly [17]. Performing evaluation has various sub-genres, we believe that load testing and configuration testing would be ideal to test the performance of our system.

### A. Load Testing

It is conducted to understand the behavior of the application under a specific expected load [16]. This load can be expected concurrent number of users on the application performing a specific number of transactions with in the set duration. This test will give out the response times of all the important business critical transactions and also point towards any bottle neck in the system.

### B. Configuration Testing

It is a variation of traditional performance testing. Rather than testing from the perspective load you are testing the effects of configuration changes in the application performance behavior.

There are seven phases of carrying out performance evaluation: identify the test environment, identify performance acceptance criteria, plan and design tests, configure the test environment, implement the test design, execute the test and analyze results and tune.

- Identify the test environment – identify the physical test environment and the tools and resources available to the test team.
- Identify performance acceptance criteria – identify the response time, throughput and resource utilization goals and constraints. Response time is a user concern, throughput is a business concern and resource utilization is a system concern.
- Plan and design tests – identify key scenarios, determine variability among representative users and how to simulate the variability, define test data and establish metrics to be collected.
- Configure the test Environment – prepare the test environment, tools and resources necessary to execute each strategy as features ad components become available for test.
- Implement the test design – develop the performance test in accordance with the test design.
- Execute the test – run and monitor your tests. Validate the tests, test data and results collection, and execute validated tests for analysis while monitoring the tests and test environment.
- Analyze Results and Tune – analyze consolidate and share result data.

The test was set up in the computing lab at FNU which allowed us to use FNU network. There is no industry standard for web application performance, in such an absence we depended on our won judgment how fast is fast enough for our application [14].We designed a set of tasks to be completed in order to measure the performance of the system.

The tests were performed using computers that have minimum hardware and software requirements to run our designed system. Software tool used to simulate task and create number of virtual users where necessary. Data was collected for the response time and was later analyzed.

Firstly we try to test the response time for our system against the existing student management system (PREMIUM). The features which are similarly available in both the systems were tested for response time. The tests were taken for eight different business critical tasks such as login, enrollment, course adjustments, generating class list, student details etc. where the features are quite similar. The result below is given for comparative test of our system against the premium system currently used by FNU.

Table IV. Results of Comparative Testing

| Task | Response delay time with FNU-CIS (ms) | Response delay time with PREMIUM (ms) |
|---|---|---|
| 1. Login | 3960 | 8018 |
| 2. Enrollment | 1637 | 4002 |
| 3. Generate Class list | 2232 | 12281 |
| 4. Course Adjustment | 2386 | 3437 |
| 5. Retrieve Coursework | 1669 | 3669 |
| 6. Student Details | 1967 | 9007 |
| 7. Student Grades | 1669 | 7950 |
| 8. Update Coursework | 1372 | 3867 |

The results of all eight tests favored FNU-CIS, thus it can be stated that it is much faster then existing systems used by FNU.

Secondly we performed load testing for our system using web load testing tool Apache JMETER. Using the software tool we simulated the tasks users will be performing on our system and tried this out with 500 virtual users which is the maximum expected load of our system at any given time. These tests were carried out on two different types of client machines. Configuration 1 had the minimum hardware and software requirements (256 MB RAM 1.2 GHZ processor) where as configuration 2 were PC's with 1 GB RAM and Intel dual core processors.

Table V. Results of Load Tesing

| Task | Avg. response delay time for Configuration 1 (ms) | Avg. response delay time for Configuration 2 (ms) |
|---|---|---|
| Login | 4663 | 4889 |
| Enrollment | 4221 | 4332 |
| Course Adjustment | 3998 | 4116 |
| Retrieve Coursework | 3774 | 4223 |
| Student Details | 2889 | 3556 |
| Class List | 2204 | 3365 |

The result of load testing verifies that there is not much time variance for the increase load with initial results of comparison testing. It also shows that there is also not much difference in the response time for running the application using different client machines.

## VI. ACKNOWLEDGMENT

I would like to thank Dr. Sharlene Dai, senior lecturer at University of the South Pacific for supervising this research project and Fiji National University for financial support to carry out this research.

## VII. CONCLUSIONS

In this paper we described three kinds of experiments done on our FNU-CIS to assess user interface, usability and performance of our system. Through heuristic evaluation, a set of problems were found and rectified. Usability and performance evaluation were based on comparative study of two systems. The results mainly favored FNU-CIS, users are very much satisfied with use of FNU-CIS and have indicated for wide use therefore we can justify that FNU-CIS is highly suitable to be used at Fiji National University

## VIII. REFERENCES

[1] Hallikainen P and Chen L (2005) "A Holistic Framework on Information Systems Evaluation with a Case Analysis", The Electronic Journal of Information Systems Evaluation, vol. 9 issue 2 pp. 57-64.

[2] Gemmell M and Pagano R (2003) "A Post Implementation Evaluation of a Student Information System in the UK Higher Education Sector", The Electronic Journal of Information Systems Evaluation, vol. 6 issue 2 pp. 95-106.

[3] Kumar B (2011) "Thin Client Web-Based Campus Information Systems for Fiji National University", International Journal of Software Engineering and Applications, vol.2, no.1 pp. 13 -26.

[4] Dai X and Kumar B (2010) "Comparing and Contrasting Campus Information Systems in South Pacific Regional Universities", 2010 International Conference on Computational and Information Sciences (ICCIS 2010), published by IEEE Computer Society pp. 721-724.

[5] Chaudhary K, Dai X and Grundy K (2010) "Experiences in Developing a Micro-payment System for Peer-to-Peer Networks", International Journal of Information Technology and Web Engineering (IJITWE), vol. 5 issue 1, pp. 23-42.

[6] Carcary M (2009) "ICT Evaluation in the Irish Higher Education Sector",The Electronic Journal of Information Systems Evaluation, vol. 12, issue 2 pp. 129-140.

[7] Alshamari M and Mayhew P (2011) "Technical Review: Current Issues of Usabilty Testing", IETE Technical Review, vol. 26, issue 6 pp. 402-406.

[8] Nielsen J and Molich R (1990) "Heuristic Evaluation of User Interfaces" CHI' 90 Proceedings ACM pp 249-256.

[9] The Basics of conduction a usabilty evaluiation, Melanie Wright. Duke University Medical Center March 26 2004.

[10] Gaffeney G "What is Uasbilty Testing?" 1999. Information and Design.

[11] S Barber (2010), How fast does a website need to Be?, unpublished.

[12] Nielsen J "How to Conduct a Heuristic Evaluation" http://www.useit.com/papers/heuristic/heuristic_evaluation.html (2011).

[13] (2011) Heuristic Evaluation – a Step by Step Guide, http://article.stiepoint.com/print/heuristic-evlautaion guide.html.

[14] (2011) How to Conduct a Heuristic Evaluation. Jakob Neilsen. http://www.useit.com/papers/heuristic_evlautaion.html.

[15] (2011) Heusrictic Evaluation, http://www.usabiltynet.org/tools/expertheuristic.html.

[16] (2011) High Performance Testing, http://www.logigear.com/ 333-highperformance-testing.html

[17] (2011) Software Performnace Testing, http://en.wikipedia.org/software_performnace_testing.